\documentclass[prl,twocolumn,superscriptaddress]{revtex4-1}

\usepackage{amsfonts,amsmath,amssymb,graphicx}

\newcommand{\ket}[1]{|#1\rangle}
\newcommand{\bra}[1]{\langle #1|}

\newcommand{\eff}{\mathrm{eff}}

\begin{document}

\title{Remote entanglement stabilization and distillation by quantum reservoir engineering}

\author{Nicolas Didier}
\altaffiliation{Current address: Rigetti Computing, 775 Heinz Avenue, Berkeley, California 94710, USA.}
\affiliation{QUANTIC team, Inria Paris, 2 rue Simone Iff, 75012 Paris, France}
\author{J\'er\'emie Guillaud}
\affiliation{QUANTIC team, Inria Paris, 2 rue Simone Iff, 75012 Paris, France}
\author{Shyam Shankar}
\affiliation{Departments of Applied Physics, Yale University, New Haven, Connecticut 06520, USA}
\author{Mazyar Mirrahimi}
\affiliation{QUANTIC team, Inria Paris, 2 rue Simone Iff, 75012 Paris, France}
\affiliation{Yale Quantum Institute, Yale University, New Haven, Connecticut 06520, USA}

\begin{abstract}
Quantum information processing in a modular architecture requires the distribution, stabilization and distillation of entanglement in a qubit network. 
We present autonomous entanglement stabilization protocols between two superconducting qubits that are coupled to distant cavities. 
The coupling between cavities is mediated and controlled via a three-wave mixing device that generates either a two-mode squeezed state or a delocalized mode between the remote cavities depending on the pump applied to the mixer. 
Local drives on the qubits and the cavities steer and maintain the system to the desired qubit Bell state. 
Most spectacularly, even a weakly-squeezed state can stabilize a maximally entangled Bell state of two distant qubits through an autonomous distillation process. 
Moreover, we show that such reservoir-engineering based protocols can stabilize entanglement in presence of qubit-cavity asymmetries and losses. 
\end{abstract}

\maketitle

{\it Introduction --}
A promising architecture for scaling up quantum machines is modular quantum computing~\cite{Monroe_2014,Monroe_2016}. An elementary task for this architecture is to entangle distant modules~\cite{Gottesman_1999,Eisert_2000,Jiang_2007}. More precisely, entangled states of remote, non-interacting qubits need to be prepared and protected against decoherence, in order that they may be available when required for quantum state transfer or gates between modules. While entangled states of remote qubits have been prepared in various quantum information platforms (e.g.~\cite{Roch_2014}), an essential but significantly harder challenge is remote entanglement stabilization. Stabilization, in this context, implies the preparation \textit{and protection} of a desired quantum state, which requires continuous correction of decoherence-induced errors. Such correction is usually achieved by some kind of feedback mechanism.

Reservoir engineering~\cite{Poyatos_1996,Verstraete_2009} is one such feedback mechanism, where the entropy produced by errors is autonomously evacuated through an engineered interaction of the system with a cold bath. Recently, reservoir engineering has been used to stabilize entanglement between two trapped ion~\cite{Wineland} and superconducting qubits~\cite{Shankar_2013, Kimchi-Schwartz_2016}. In these schemes, the entangled qubits share a common dissipative mode consisting of a motional degree of freedom or a common resonant cavity, respectively. In order to extend these schemes from local to distant modules, nonlocal correlations need to be generated between the two modules containing the qubits. Moreover, these correlations must be generated in a manner that is amenable to scaling up to multiple modules connected through a quantum router~\cite{Monroe_2014}.

\begin{figure}[t]
\includegraphics[width=\columnwidth]{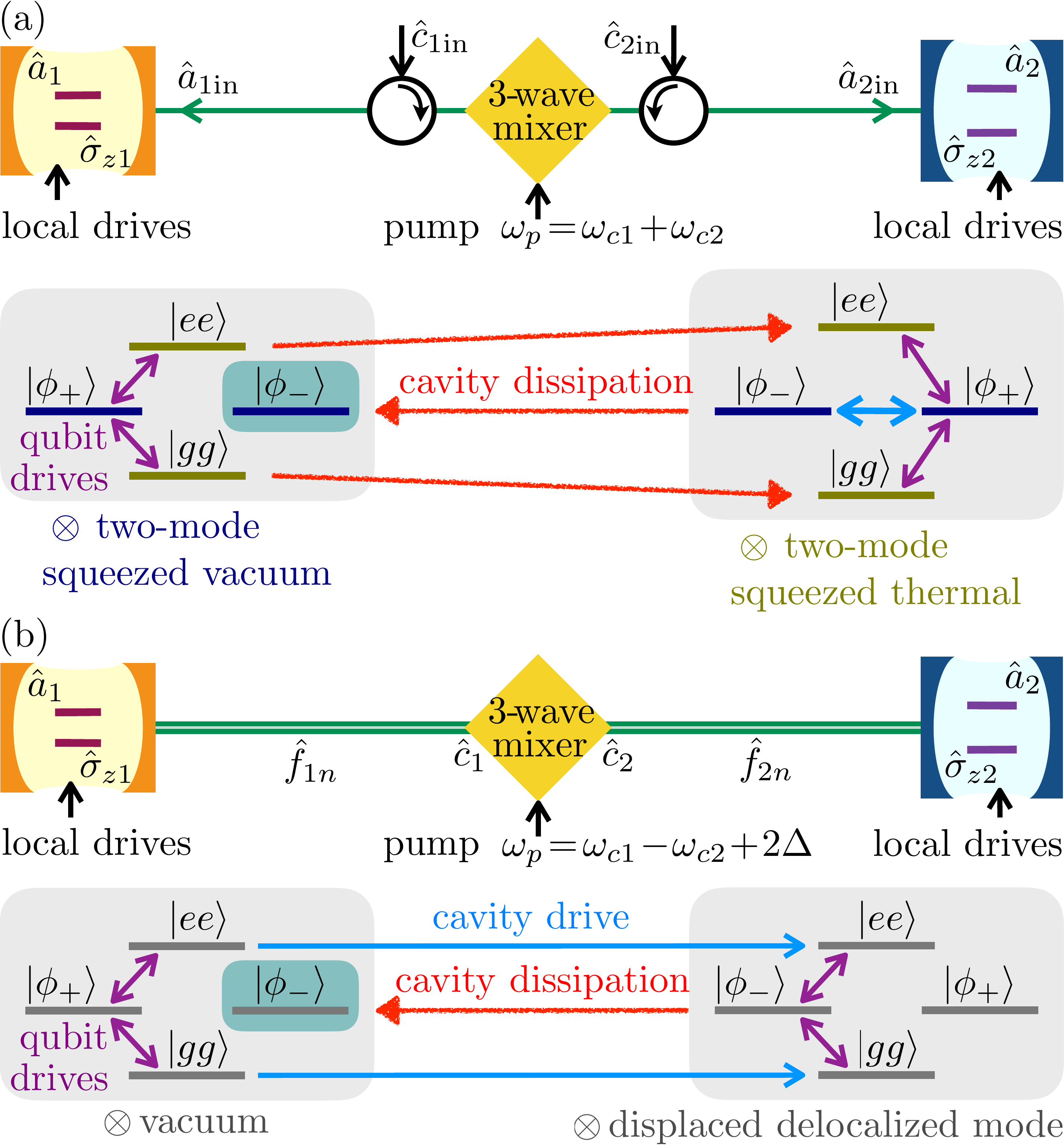}
\caption{Autonomous remote entanglement stabilization protocols with three-wave mixers.
(a)~For a directional coupling in the amplification mode, the TWM is a source of two-mode squeezed light that steers the cavities to two-mode squeezed vacuum for the odd qubit subspace and to a two-mode squeezed thermal state for the even qubit subspace (red arrows). Qubit drives couple the two subspaces (purple arrows). A coupling between the Bell state $\ket{\phi_+}$ and $\ket{\phi_-}$ is activated by the fluctuations of the photon number imbalance present in the two-mode squeezed thermal state (light blue arrow). Entanglement accumulation leads to the stabilization of the Bell state $\ket{\phi_-}$ in two-mode squeezed vacuum with $100\,\%$ fidelity.
(b)~For a bidirectional coupling in the conversion mode, the TWM generates delocalized modes between the two distant cavities. 
These delocalized modes are displaced by the cavity drives for even qubit states (light blue arrows). Qubit drives couple the even and odd subspaces (purple arrows). Cavity dissipation steers the cavities to vacuum for the odd qubit subspace (red arrow).
For large enough coupling the physics becomes effectively single mode and the protocol of Ref.~\onlinecite{Shankar_2013} can be applied.}
\label{figscheme}
\end{figure}

In this paper we propose two reservoir engineering protocols for stabilizing entanglement between distant qubits. Both protocols use a three-wave mixer (TWM) to couple distant cavity modes that contain the qubits. We specialize to the case of superconducting quantum circuits where three-wave mixing is realized with a Josephson Parametric Converter~\cite{Bergeal_2010}, a device routinely used in experiments. This device controls the interaction between a pair of field modes by pumping a third one; the nature and strength of the coupling is set by the pump frequency and amplitude, respectively.
The TWM is a versatile device, that can switch an interaction with an on/off ratio in excess of $10^4$. Moreover, it can perform two operations --- amplification with a two-mode squeezing interaction~\cite{Flurin_2012,Flurin_2015}, or frequency conversion with a beam-splitter interaction~\cite{Abdo_2013b,Sirois_2015}.
There are furthermore two kinds of connection between the distant cavites and the TWM, unidirectional coupling via circulators or bidirectional coupling via long transmission lines~\cite{Sundaresan_2015}, as depicted in Fig.~\ref{figscheme}.
The two remote entanglement stabilization schemes we describe in this paper correspond to two different combinations of operation and connection between the cavities and the TWM.

The first scheme for entanglement stabilization can be understood as an autonomous distillation protocol~\cite{Bennett_1996,Vollbrecht_2011}. It uses a two-mode squeezing interaction, generated by the TWM for a pump at the frequency sum of the two other TWM modes, which are assumed to be resonant with the distant cavities. The interaction continuously injects an entangled two-mode squeezed state into the distant cavities, as in Fig.~\ref{figscheme}~(a). Drives at qubit transitions together with cavity dissipation continuously distill entanglement out of this two-mode squeezed state into a non-local qubit Bell state. Our scheme achieves $100\,\%$ fidelity \textit{even with weak squeezing}. This property makes it fundamentally different from a previous proposal~\cite{Kraus_2004}, that requires squeezing in resonance with the qubit frequencies and reaches high fidelity only with strong squeezing. While Ref.~\cite{Kraus_2004} illustrates entanglement transfer between flying field modes and qubits, our protocol uses   cavities as ancillary stationary field modes, and distills entanglement out of them to stabilize a maximally entangled state of the qubits. Moreover, in contrast to~\cite{Vollbrecht_2011}, our protocol is based on concrete dispersive interaction Hamiltonians that are routinely used in various physical platforms such as superconducting circuits.

Our second entanglement stabilization scheme generalizes the protocol of Ref.~\cite{Shankar_2013} to the case of distant qubits. It employs a beam-splitter interaction, which is obtained when the TWM pump frequency is at the frequency difference of the two other TWM modes.
When the distant cavities are coupled to the TWM through long transmission lines, as in Fig.~\ref{figscheme}~(b), we show that they act effectively as a single delocalized mode. The stabilization protocol of Ref.~\cite{Shankar_2013} can then be applied to this delocalized mode to stabilize a nonlocal qubit Bell state.
Interestingly, the crucial dispersive shift symmetry of the two qubits required in the local protocol~\cite{Shankar_2013} is lifted by tuning the TWM pump detuning. This feature distinguishes our protocol from~\cite{Aron_2014,Kimchi-Schwartz_2016} that requires a symmetry on the qubits and the cavities combined with close enough cavities to get strong tunnel coupling. Furthermore, the bidirectional nature of the coupling offers better protection against transmission losses~\cite{motzoi_2016,Shi_2015}, akin to population transfer in optomechanics using dark modes~\cite{Wang_2012}. We describe both protocols in detail in the following and discuss their robustness against imperfections.

{\it Two-mode squeezed states --}
The TWM couples the three modes $\hat{c}_1$, $\hat{c}_2$ and $\hat{c}_3$ according to the Hamiltonian
$\hat{H}_\mathrm{TWM}=\hbar g_3(\hat{c}_1^\dag+\hat{c}_1)(\hat{c}_2^\dag+\hat{c}_2)(\hat{c}_3^\dag+\hat{c}_3)$  ($g_3$ denoting the coupling strength),
where the third mode is strongly driven, it plays the role of the pump and is treated classically.
Amplification is obtained for a pump frequency~$\omega_p$ set to the sum $\omega_{c1}+\omega_{c2}$ and
the TWM becomes a two-mode squeezer.
For directional coupling, obtained with circulators as sketched in Fig.~\ref{figscheme}~(a),
the TWM acts as a correlated bath for the distant cavities.
Their dissipative dynamics is governed by the Lindbladian~\cite{Kraus_2004}
\begin{align}
\hat{L}_\mathrm{S}=\kappa D[\hat{a}_1\cosh r+\hat{a}_2^\dag\sinh r]+\kappa D[\hat{a}_2\cosh r+\hat{a}_1^\dag\sinh r]
\label{lindbladian}
\end{align}
with the dissipation superoperator $D[\hat{a}]\cdot=\hat{a}\cdot\hat{a}^\dag-\frac{1}{2}\{\hat{a}^\dag\hat{a},\cdot\}$. Here $\kappa$ denotes the cavities decay rates assumed to be identical, and the squeezing parameter $r$ is set by the pump amplitude, also expressed in decibel: 
$r_\mathrm{dB}=(20/\log10)r$~\cite{Note1}. For empty cavities, this Lindbladian steers the cavity state to the two-mode squeezed vacuum state
$\hat{\rho}_\mathrm{SV}=\hat{S}_r\ket{0,0}\bra{0,0}\hat{S}_r^\dag$, 
with the two-mode squeezing operator
$\hat{S}_r=e^{r(\hat{a}_1\hat{a}_2-\hat{a}_1^\dag\hat{a}_2^\dag)}$.

The qubits are dispersively coupled to their own cavity,
with the coupling Hamiltonian, $\hat{H}_{\mathrm{dispersive},j=1,2}=-\tfrac{1}{2}\hbar\chi_j\hat{a}_j^\dag\hat{a}_j\hat{\sigma}_{zj}$ ($\chi_j$ representing the qubit-cavity dispersive coupling strengths).
Considering equal dispersive shifts, the qubit-cavity Hamiltonian reads
\begin{align}
\hat{H}_\mathrm{dispersive}=\tfrac{1}{2}\hbar\chi\hat{N}(\hat{\sigma}_\mathrm{gg}-\hat{\sigma}_\mathrm{ee})+\tfrac{1}{2}\hbar\chi\hat{M}(\hat{\sigma}_{-+}+\hat{\sigma}_{+-}),
\label{Hdispersive}
\end{align}
with $\hat{\sigma}_{kl}=\ket{k}\bra{l}$, $\ket{\mathrm{g}}=\ket{gg}$, $\ket{\mathrm{e}}=\ket{ee}$
and $\hat{\sigma}_{-+}$, $\hat{\sigma}_{+-}$ couple the odd parity Bell states $\ket{\phi_\mp}=\frac{1}{\sqrt{2}}(\ket{ge}\mp\ket{eg})$.
The even and odd qubit subspaces are coupled to the photon number sum $\hat{N}=\hat{a}_1^\dag\hat{a}_1+\hat{a}_2^\dag\hat{a}_2$ and difference $\hat{M}=\hat{a}_1^\dag\hat{a}_1-\hat{a}_2^\dag\hat{a}_2$.
Similarly to~\cite{Shankar_2013}, the qubits are driven at resonance with the same Rabi amplitudes $\Omega$, leading to a Hamiltonian of the form
$\hat{H}_\mathrm{drive}=\sqrt{2}\hbar\Omega(\hat{\sigma}_{+\mathrm{g}}+\hat{\sigma}_{+\mathrm{e}})+\mathrm{h.c.}$,
coupling the even qubit subspace to $\ket{\phi_+}$.

To highlight the stabilization mechanism of $\ket{\phi_-}$ by entanglement distillation, let us have a look at the quantum dynamics 
sketched in Fig.~\ref{figscheme}~(a).
Whenever the qubits are in the even manifold Span$\{\ket{gg},\ket{ee}\}$, the correlated dissipation, Eq.~\eqref{lindbladian}, combined with the dispersive coupling, Eq.~\eqref{Hdispersive}, generate a two-mode squeezed thermal state $\hat{\rho}_\mathrm{ST}$ (defined in~\cite{SM}), the thermal aspect comes from the detuning of the cavities by $\chi/2$.
Qubit driving then induces oscillations between Span$\{\ket{gg},\ket{ee}\}$ and $\ket{\phi_+}$. Finally, 
the Bell state $\ket{\phi_-}$ is populated from $\ket{\phi_+}$ via the dispersive interaction.
It is essential here to notice that the coupling is mediated by the photon number imbalance $\hat{M}$,
the thermal character of $\hat{\rho}_\mathrm{ST}$ is thus crucial to activate this tunneling process.
Once in the odd qubit subspace, the Lindbladian steers the cavities to the two-mode squeezed vacuum state $\hat{\rho}_\mathrm{SV}$~\cite{SM}.
At each round, entanglement is accumulated and 
the cycle in the Hilbert space stops when in $\ket{\phi_-}\bra{\phi_-}\hat{\rho}_\mathrm{SV}$.

\begin{figure}[t]
\includegraphics[width=\columnwidth]{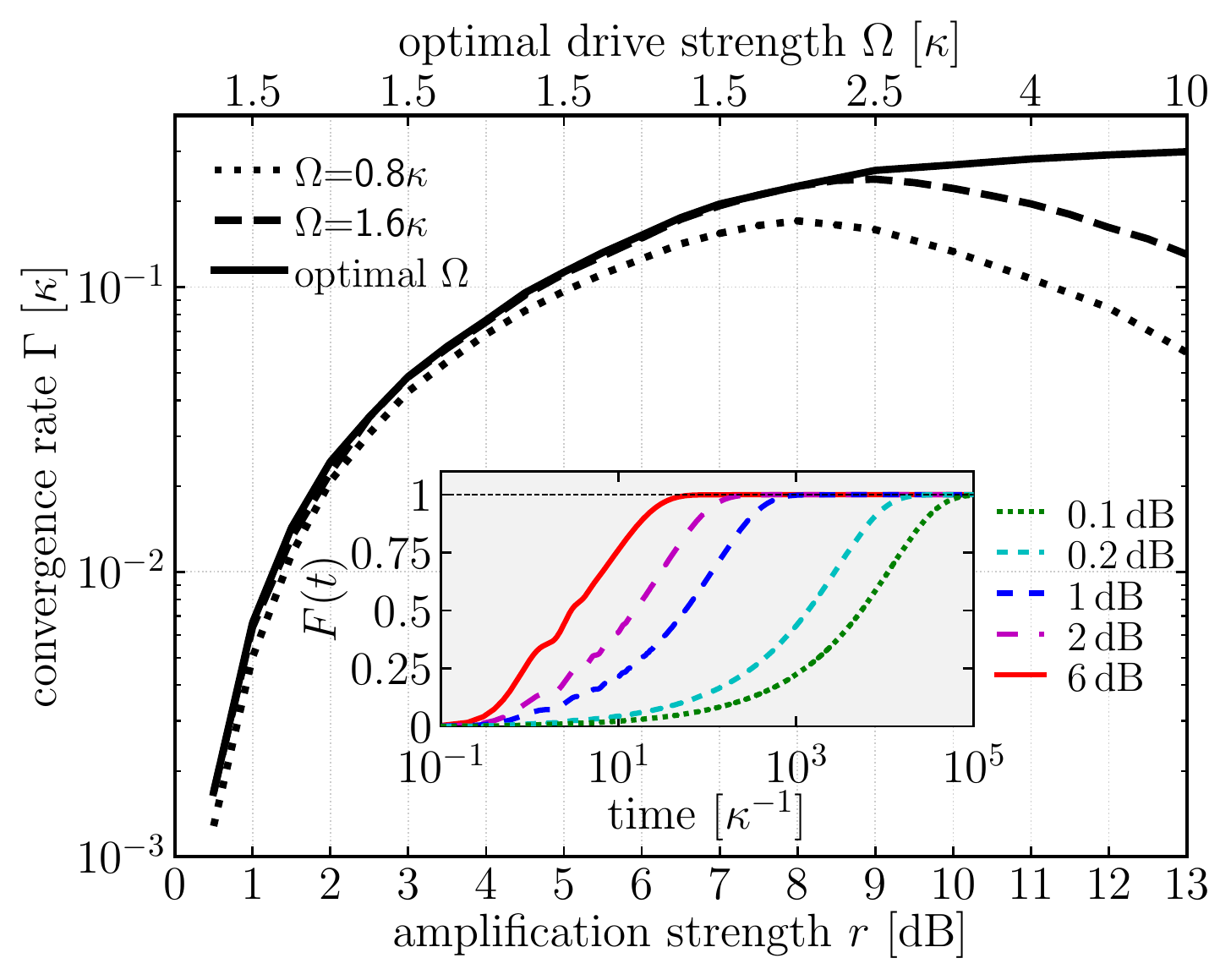}
\caption{Remote entanglement stabilization in the amplification mode and with the directional connection.
The convergence rate is plotted versus amplification strength in dB. For a fixed strength $\Omega$ of the qubit drives, the convergence rate reaches a maximum for a finite value of the squeezing strength. However this decrease  in convergence rate, observed for strong squeezing powers, can be compensated by increasing the qubit drives strength (full curve, corresponding to an optimal choice of $\Omega$).
Parameters are $\chi_1/2\pi=\chi_2/2\pi=5\,\mathrm{MHz}$, $\kappa/2\pi=1\,\mathrm{MHz}$.
Inset: Temporal dynamics of the fidelity $F$ for different values of the squeezing strength.}
\label{figamplification}
\end{figure}

The convergence rate $\Gamma$ to the Bell state $\ket{\phi_-}$ is plotted in Fig.~\ref{figamplification} versus squeezing strength.
For a fixed Rabi rate $\Omega=0.8\kappa$, the highest rate takes place at a modest squeezing strength, below $8\,$dB.
This behavior can be explained by two competing phenomena~\cite{SM} that can be compensated by increasing the Rabi rate, see top axis, leading to a saturation around $\Gamma_{\mathrm{opt}} \simeq \kappa/3$.

\begin{figure}[t]
\includegraphics[width=\columnwidth]{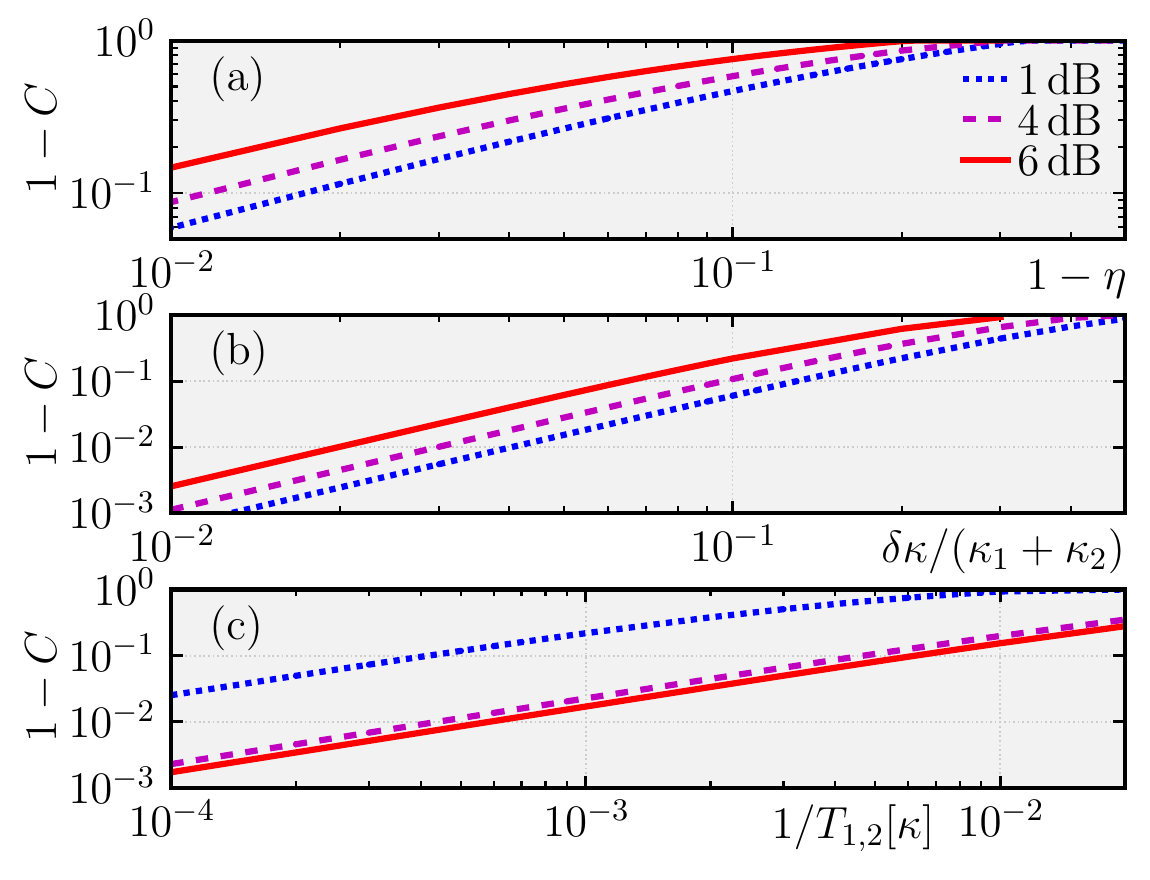}
\caption{Effect of imperfections on remote entanglement stabilization in the amplification mode and directional connection.
Using the same parameters as in Fig.~\ref{figamplification}, and $\Omega = 1.6\kappa$, the complementary of the qubits concurrence is plotted against 
transmission losses~(a),  asymmetry in the cavity couplings to the transmission lines~(b) and qubit's relaxation and dephasing (for $T_1=T_2$)~(c). One observes that applying a weak amount of  squeezing provides more robustness to transmission losses and asymmetries but leads to a slower rate of convergence, harming the asymptotic concurrence in presence of qubit's relaxation and dephasing. In practice, one needs to choose an optimal value of squeezing to compromise between these two effects.}
\label{figamplificationimperfection}
\end{figure}

The effect of photon losses in the transmission lines is to heat the two-mode squeezed state for the odd qubit subspace.
As a result, the two-mode squeezed vacuum state is not the steady-state and cannot stabilize the Bell state with $100\,\%$ fidelity anymore.
In  Fig.~\ref{figamplificationimperfection}~(a), we plot the complementary of the qubits concurrence $C$~\cite{Wootters_2001}. The concurrence decreases with the losses (here, $\eta$ represents the transmission efficiency between the TWM and each of the cavities).
This situation is similar to~\cite{Kraus_2004} but with the crucial difference that with no loss our protocol gives a fidelity of $100\,\%$ and that a smaller amount of squeezing provides further robustness to  transmission losses. 
The  asymmetry in the cavity couplings to the transmission lines ($\delta\kappa=|\kappa_1-\kappa_2|)$ also reduces the steady-state concurrence, but again, smaller amount of squeezing leads to further robustness, see Fig.~\ref{figamplificationimperfection}~(b).
The results obtained so far did not take into account the effect of qubit relaxation (rate $1/T_1$) and dephasing (rate $1/T_2$).
Naturally, the steady-state concurrence decreases for large relaxation and dephasing rates, see Fig.~\ref{figamplificationimperfection}~(c).
While a larger amount of squeezing provides a faster entanglement stabilization, therefore compensating for qubits relaxation and dephasing,  it reduces the robustness to transmission loss and asymmetry. The optimal value of squeezing should be chosen as a function of the experimental parameters. 
Our protocol can also be realized between field modes at identical frequencies~\cite{Roy_2016}, such a setup may be helpful in the modular architecture context because the cavity modes in distant modules could have identical frequency. This is accomplished by creating single-mode squeezed states with a pair of degenerate parametric amplifiers which are then transformed into a two-mode squeezed state by a 50-50 beam splitter.

{\it Delocalized mode --}
In the conversion mode, the pump tone frequency $\omega_p$ is set close to the difference $\omega_{c1}-\omega_{c2}$ between the frequency of the modes $\hat{c}_1$ and $\hat{c}_2$.
The TWM then generates a beam-splitter interaction,
\begin{align}
\hat{H}_\mathrm{convert}=\hbar\Delta(-\hat{c}_1^\dag\hat{c}_1+\hat{c}_2^\dag\hat{c}_2)+\hbar g(\hat{c}_1^\dag\hat{c}_2+\hat{c}_2^\dag\hat{c}_1),
\label{Hconvert}
\end{align}
in the appropriate rotating frame and after a rotating wave approximation.
The detuning,
$\Delta=\frac{1}{2}(\omega_p-\omega_{c1}+\omega_{c2})$,
appears to be very useful to correct mismatch between the qubit-cavity couplings.

\begin{figure}[t]
\includegraphics[width=\columnwidth]{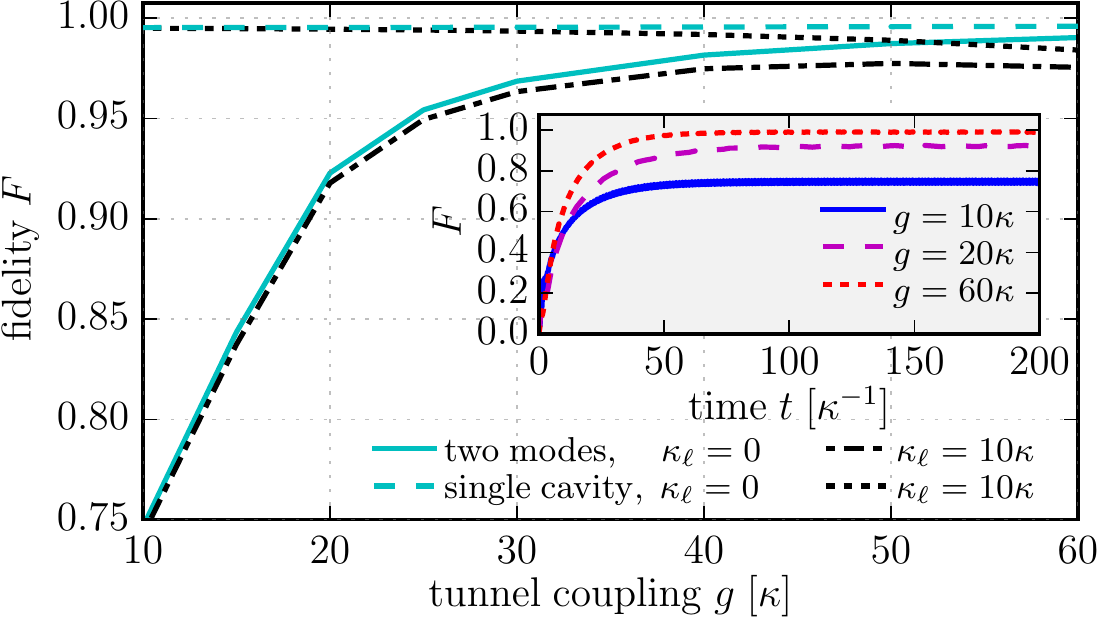}
\caption{Remote entanglement stabilization in the conversion mode and with the bidirectional connection.
The single-cavity result (dashed and dotted lines) is recovered for a sufficiently large tunnel coupling~$g$ (full and dash-dotted lines).
Results are obtained for different values of the damping rate of the long resonators, $\kappa_{\ell}=0$ in cyan and $\kappa_{\ell}=10\kappa$ in black.
The damping rate of the cavities and TWM modes is $\kappa/2\pi=1\,\mathrm{MHz}$,
the effective dispersive shift is equal to $\chi_\eff=5\kappa$,
the cavity drives are set to displace the mode $\hat{d}_1$ by an average of $\bar{n}=3$ photons,
the qubit drives strength is $\Omega=\kappa_\eff/2$.
For the simulations of the single-cavity case, we vary the effective damping rate $\kappa_\eff$ similarly to the two-mode case. This explains the slight reduction of the fidelity for $\kappa_{\ell}=10\kappa$ with increasing $g$ (and therefore $\kappa_\eff$).
Inset: Temporal dynamics of fidelity establishment and stabilization for different values of the tunnel coupling and $\kappa_{\ell}=\kappa$.}
\label{figconversion}
\end{figure}

To expose the remote connection idea, as a toy model we first consider a full hybridization between the cavity mode $\hat{a}_1$ ($\hat{a}_2$) with the TWM mode $\hat{c}_1$ ($\hat{c}_2$), i.e.~$\hat{c}\to\hat{a}$ in~\eqref{Hconvert}.
Hamiltonian~\eqref{Hconvert} is diagonalized through rotating the modes $\hat{a}_j$ by an angle defined as $\tan2\theta=g/\Delta$, 
$\begin{pmatrix}\hat{d}_1&\hat{d}_2\end{pmatrix}^\intercal=R_\theta\begin{pmatrix}\hat{a}_1&\hat{a}_2\end{pmatrix}^\intercal$.
In order to couple the mode $\hat{d}_1$ to both qubits with the same dispersive shift $\chi_\eff=\frac{\chi_1\chi_2}{\chi_1+\chi_2}$,
the detuning is set to 
$\Delta=g\frac{\chi_1-\chi_2}{2\sqrt{\chi_1\chi_2}}$.
The total Hamiltonian is then given by
\begin{align}
\hat{H}_\mathrm{toy}&=\tfrac{1}{2}E(\hat{d}_1^\dag\hat{d}_1-\hat{d}_2^\dag\hat{d}_2)-\tfrac{1}{2}\hbar\chi_\eff(\hat{\sigma}_{z1}+\hat{\sigma}_{z2})\hat{d}_1^\dag\hat{d}_1,
\label{Htoy}
\end{align}
with the energy separation $E=\hbar g\frac{\chi_1+\chi_2}{\sqrt{\chi_1\chi_2}}$.
For a tunnel coupling $g$ strong enough to ensure $E\gg \hbar\chi_\eff$, the delocalized mode $\hat{d}_2$ is largely detuned from $\hat{d}_1$.
As a consequence, we have neglected the coupling between $\hat{d}_1$ and $\hat{d}_2$ through the dispersive interaction.
The physics becomes effectively single mode and the protocol experimentally realized in~\cite{Shankar_2013} can be applied to the delocalized mode~$\hat{d}_1$.
Manipulating mode~$\hat{d}_1$ is straightforwardly achieved with drives on one or both cavities.
The final Hamiltonian after a rotating wave approximation in the large tunnel coupling $g$ limit in the interaction picture reads
\begin{align}
\hat{H}_\eff=&\chi_\eff(\hat{\sigma}_\mathrm{gg}-\hat{\sigma}_\mathrm{ee})\hat{d}_1^\dag\hat{d}_1
+\kappa\sqrt{\bar{n}}\cos(\chi_\eff t)(\hat{d}_1^\dag+\hat{d}_1)\nonumber\\
+\{\Omega&(\hat{\sigma}_{+\mathrm{g}}+\hat{\sigma}_{\mathrm{e}+})-\Omega e^{i\bar{n}\chi_\eff t}(\hat{\sigma}_{-\mathrm{g}}-\hat{\sigma}_{\mathrm{e}-})+\mathrm{h.c.}\}.
\label{Heff}
\end{align}
The first term of Hamiltonian $\hat{H}_\eff$ shifts the frequency of the cavities by $\pm\chi_\eff$ if the qubits are both in ground or excited state,
but does not involve the qubit states $\ket{ge}$ and $\ket{eg}$ that contribute to the Bell state to be stabilized.
This crucial property is the result of correcting the dispersive shift asymmetry with $\Delta$.
The second term displaces the cavity states to a coherent state with an average of $\bar{n}$ photons if the qubits are in states $\ket{gg}$ or $\ket{ee}$,
it corresponds to the cavity drive channel in Fig.~\ref{figscheme}~(b).
The last terms drive the qubit to the Bell state $\ket{\phi_+}$ if the cavities are empty and to the Bell state $\ket{\phi_-}$ if the cavities are displaced, respectively.
Both cavities are also coupled to zero-temperature environnements with damping rate~$\kappa$,
that tends to relaxe the cavities to vacuum.
The resulting quantum dynamics stabilizes the state $\ket{\phi_-}\ket{00}$.
The fidelity $F$ of the Bell state $\ket{\phi_-}$ is plotted in Fig.~\ref{figconversion} versus tunnel coupling strength and compared to the result obtained for two qubits in the same cavity~\cite{Shankar_2013}.
The remote entanglement stabilization converges to the local protocol result for $g\gtrsim40\kappa$.
In this limit, the characteristics of the stabilization process have been widely studied for a single cavity~\cite{Leghtas_2013,Liu_2016} and can be directly used to describe the remote configuration.
Temporal dynamics for the preparation and stabilization protocol are plotted in the inset of Fig.~\ref{figconversion}.

We now consider the cavities coupled to the TWM through long resonators of fundamental frequency well below the cavity and TWM frequencies~\cite{Sundaresan_2015}.
Among their harmonics we select the closest to each cavity frequency, $\hat{f}_1$ and $\hat{f}_2$, coupled to the cavity (TWM) with strength $\tilde{g}\sin\varphi$ ($\tilde{g}\cos\varphi$).
Very interestingly, in the limit of large coupling to the long resonators, $\tilde{g}\gg g,\Delta$, the dynamics can be restricted to the two delocalized modes $\hat{d}_1$, $\hat{d}_2$ involving the cavity and TWM modes with a vanishing contribution of the connecting modes.
The modes $\hat{d}_{1,2}$ are thus delocalized dark modes with respect to the long resonators.
The resulting effective dispersive coupling is reduced~\cite{SM},
the extension to remote cavities thus requires strong dispersive couplings, $\chi_{1,2}\sim20\kappa$.

A more rigorous description of the whole system, involving the harmonics $\hat{f}_{1n}$, $\hat{f}_{2n}$ of the long resonators, shows that the effective single mode description is valid for large enough tunnel coupling $g$ and away from other resonances~\cite{SM}.
In Fig.~\ref{figconversion} the parameters of the delocalized modes $\hat{d}_1$ and $\hat{d}_2$ are obtained after diagonalization of the linear system.
The parameters are $\omega_{c1}/2\pi=7\,\mathrm{GHz}$, $\omega_{c2}/2\pi=11\,\mathrm{GHz}$,
$\chi_1/2\pi=18.8\,\mathrm{MHz}$ and $\chi_2/2\pi=15\,\mathrm{MHz}$,
$\tilde{g}/2\pi=100\,\mathrm{MHz}$,
$\varphi=0.22\pi$.
The fundamental frequency of the long resonators is $\omega_{f0}\simeq1\,\mathrm{GHz}$,
the detuning between the cavities and the TWM modes is set to $\pm10\,\mathrm{MHz}$,
as for the detuning between the cavity mode $\hat{a}_1$ ($\hat{a}_2$) and the $7^\mathrm{th}$ ($11^\mathrm{th}$) harmonic $\hat{f}_{1,7}$ ($\hat{f}_{2,11}$).
Concerning the robustness to imperfections,
the pump detuning is used to correct a dispersive coupling asymmetry and
an asymmetry in other parameters does not lead to qualitative changes in performances.
Only a large damping rate of the connecting resonators, $\kappa_{\ell}=10\kappa$, can affect the stabilized fidelity (see Fig.~\ref{figconversion}).

{\it Conclusion --}
We have proposed two protocols for entanglement stabilization of two distant qubits that are connected to a three-wave mixer.
The stabilization mechanism is controlled by the TWM pump amplitude and frequency and use different nonlocal resources, either a two-mode squeezed state or a delocalized mode. The latter protocol is more robust against experimental imperfections such as asymmetries and losses. On the other hand, the protocol using a shared two-mode squeezed state, is an unusual example of autonomous entanglement distillation where even modest squeezing results in significant entanglement. Our protocols are general and can be implemented with current superconducting quantum circuit technology. Remote entanglement stabilization based on TWMs can be used to distribute entanglement in a network of qubits as the connection between qubits and the mixer could be through a quantum router~\cite{Monroe_2014} with no essential change in the physics of the process. Thus the schemes are well suited for modular quantum computing and demonstrate that TWMs are advantageous devices for this purpose.

\begin{acknowledgments}
{\it Acknowledgements --}
We gratefully acknowledge useful discussions with Aashish Clerk, Michel Devoret and Benjamin Huard.
This research was supported by the Agence Nationale de la Recherche under grant ANR-14-CE26-0018, by Inria's DPEI under the TAQUILLA associated team and by ARO under Grant No. W911NF-14-1-0011.
\end{acknowledgments}


\begin{thebibliography}{35}%
\makeatletter
\providecommand \@ifxundefined [1]{%
 \@ifx{#1\undefined}
}%
\providecommand \@ifnum [1]{%
 \ifnum #1\expandafter \@firstoftwo
 \else \expandafter \@secondoftwo
 \fi
}%
\providecommand \@ifx [1]{%
 \ifx #1\expandafter \@firstoftwo
 \else \expandafter \@secondoftwo
 \fi
}%
\providecommand \natexlab [1]{#1}%
\providecommand \enquote  [1]{``#1''}%
\providecommand \bibnamefont  [1]{#1}%
\providecommand \bibfnamefont [1]{#1}%
\providecommand \citenamefont [1]{#1}%
\providecommand \href@noop [0]{\@secondoftwo}%
\providecommand \href [0]{\begingroup \@sanitize@url \@href}%
\providecommand \@href[1]{\@@startlink{#1}\@@href}%
\providecommand \@@href[1]{\endgroup#1\@@endlink}%
\providecommand \@sanitize@url [0]{\catcode `\\12\catcode `\$12\catcode
  `\&12\catcode `\#12\catcode `\^12\catcode `\_12\catcode `\%12\relax}%
\providecommand \@@startlink[1]{}%
\providecommand \@@endlink[0]{}%
\providecommand \url  [0]{\begingroup\@sanitize@url \@url }%
\providecommand \@url [1]{\endgroup\@href {#1}{\urlprefix }}%
\providecommand \urlprefix  [0]{URL }%
\providecommand \Eprint [0]{\href }%
\providecommand \doibase [0]{http://dx.doi.org/}%
\providecommand \selectlanguage [0]{\@gobble}%
\providecommand \bibinfo  [0]{\@secondoftwo}%
\providecommand \bibfield  [0]{\@secondoftwo}%
\providecommand \translation [1]{[#1]}%
\providecommand \BibitemOpen [0]{}%
\providecommand \bibitemStop [0]{}%
\providecommand \bibitemNoStop [0]{.\EOS\space}%
\providecommand \EOS [0]{\spacefactor3000\relax}%
\providecommand \BibitemShut  [1]{\csname bibitem#1\endcsname}%
\let\auto@bib@innerbib\@empty
\bibitem [{\citenamefont {Monroe}\ \emph {et~al.}(2014)\citenamefont {Monroe},
  \citenamefont {Raussendorf}, \citenamefont {Ruthven}, \citenamefont {Brown},
  \citenamefont {Maunz}, \citenamefont {Duan},\ and\ \citenamefont
  {Kim}}]{Monroe_2014}%
  \BibitemOpen
  \bibfield  {author} {\bibinfo {author} {\bibfnamefont {C.}~\bibnamefont
  {Monroe}}, \bibinfo {author} {\bibfnamefont {R.}~\bibnamefont {Raussendorf}},
  \bibinfo {author} {\bibfnamefont {A.}~\bibnamefont {Ruthven}}, \bibinfo
  {author} {\bibfnamefont {K.~R.}\ \bibnamefont {Brown}}, \bibinfo {author}
  {\bibfnamefont {P.}~\bibnamefont {Maunz}}, \bibinfo {author} {\bibfnamefont
  {L.-M.}\ \bibnamefont {Duan}}, \ and\ \bibinfo {author} {\bibfnamefont
  {J.}~\bibnamefont {Kim}},\ }\href {\doibase 10.1103/PhysRevA.89.022317}
  {\bibfield  {journal} {\bibinfo  {journal} {Phys. Rev. A}\ }\textbf {\bibinfo
  {volume} {89}},\ \bibinfo {pages} {022317} (\bibinfo {year}
  {2014})}\BibitemShut {NoStop}%
\bibitem [{\citenamefont {Monroe}\ \emph {et~al.}(2016)\citenamefont {Monroe},
  \citenamefont {Schoelkopf},\ and\ \citenamefont {Lukin}}]{Monroe_2016}%
  \BibitemOpen
  \bibfield  {author} {\bibinfo {author} {\bibfnamefont {C.~R.}\ \bibnamefont
  {Monroe}}, \bibinfo {author} {\bibfnamefont {R.~J.}\ \bibnamefont
  {Schoelkopf}}, \ and\ \bibinfo {author} {\bibfnamefont {M.~D.}\ \bibnamefont
  {Lukin}},\ }\href@noop {} {\bibfield  {journal} {\bibinfo  {journal} {Sci.
  Am.}\ }\textbf {\bibinfo {volume} {314}},\ \bibinfo {pages} {50} (\bibinfo
  {year} {2016})}\BibitemShut {NoStop}%
\bibitem [{\citenamefont {Gottesman}\ and\ \citenamefont
  {Chuang}(1999)}]{Gottesman_1999}%
  \BibitemOpen
  \bibfield  {author} {\bibinfo {author} {\bibfnamefont {D.}~\bibnamefont
  {Gottesman}}\ and\ \bibinfo {author} {\bibfnamefont {I.~L.}\ \bibnamefont
  {Chuang}},\ }\href@noop {} {\bibfield  {journal} {\bibinfo  {journal}
  {Nature}\ }\textbf {\bibinfo {volume} {402}},\ \bibinfo {pages} {390}
  (\bibinfo {year} {1999})}\BibitemShut {NoStop}%
\bibitem [{\citenamefont {Eisert}\ \emph {et~al.}(2000)\citenamefont {Eisert},
  \citenamefont {Jacobs}, \citenamefont {Papadopoulos},\ and\ \citenamefont
  {Plenio}}]{Eisert_2000}%
  \BibitemOpen
  \bibfield  {author} {\bibinfo {author} {\bibfnamefont {J.}~\bibnamefont
  {Eisert}}, \bibinfo {author} {\bibfnamefont {K.}~\bibnamefont {Jacobs}},
  \bibinfo {author} {\bibfnamefont {P.}~\bibnamefont {Papadopoulos}}, \ and\
  \bibinfo {author} {\bibfnamefont {M.~B.}\ \bibnamefont {Plenio}},\ }\href
  {\doibase 10.1103/PhysRevA.62.052317} {\bibfield  {journal} {\bibinfo
  {journal} {Phys. Rev. A}\ }\textbf {\bibinfo {volume} {62}},\ \bibinfo
  {pages} {052317} (\bibinfo {year} {2000})}\BibitemShut {NoStop}%
\bibitem [{\citenamefont {Jiang}\ \emph {et~al.}(2007)\citenamefont {Jiang},
  \citenamefont {Taylor}, \citenamefont {S\o{}rensen},\ and\ \citenamefont
  {Lukin}}]{Jiang_2007}%
  \BibitemOpen
  \bibfield  {author} {\bibinfo {author} {\bibfnamefont {L.}~\bibnamefont
  {Jiang}}, \bibinfo {author} {\bibfnamefont {J.~M.}\ \bibnamefont {Taylor}},
  \bibinfo {author} {\bibfnamefont {A.~S.}\ \bibnamefont {S\o{}rensen}}, \ and\
  \bibinfo {author} {\bibfnamefont {M.~D.}\ \bibnamefont {Lukin}},\ }\href
  {\doibase 10.1103/PhysRevA.76.062323} {\bibfield  {journal} {\bibinfo
  {journal} {Phys. Rev. A}\ }\textbf {\bibinfo {volume} {76}},\ \bibinfo
  {pages} {062323} (\bibinfo {year} {2007})}\BibitemShut {NoStop}%
\bibitem [{\citenamefont {Roch}\ \emph {et~al.}(2014)\citenamefont {Roch},
  \citenamefont {Schwartz}, \citenamefont {Motzoi}, \citenamefont {Macklin},
  \citenamefont {Vijay}, \citenamefont {Eddins}, \citenamefont {Korotkov},
  \citenamefont {Whaley}, \citenamefont {Sarovar},\ and\ \citenamefont
  {Siddiqi}}]{Roch_2014}%
  \BibitemOpen
  \bibfield  {author} {\bibinfo {author} {\bibfnamefont {N.}~\bibnamefont
  {Roch}}, \bibinfo {author} {\bibfnamefont {M.~E.}\ \bibnamefont {Schwartz}},
  \bibinfo {author} {\bibfnamefont {F.}~\bibnamefont {Motzoi}}, \bibinfo
  {author} {\bibfnamefont {C.}~\bibnamefont {Macklin}}, \bibinfo {author}
  {\bibfnamefont {R.}~\bibnamefont {Vijay}}, \bibinfo {author} {\bibfnamefont
  {A.~W.}\ \bibnamefont {Eddins}}, \bibinfo {author} {\bibfnamefont {A.~N.}\
  \bibnamefont {Korotkov}}, \bibinfo {author} {\bibfnamefont {K.~B.}\
  \bibnamefont {Whaley}}, \bibinfo {author} {\bibfnamefont {M.}~\bibnamefont
  {Sarovar}}, \ and\ \bibinfo {author} {\bibfnamefont {I.}~\bibnamefont
  {Siddiqi}},\ }\href {\doibase 10.1103/PhysRevLett.112.170501} {\bibfield
  {journal} {\bibinfo  {journal} {Phys. Rev. Lett.}\ }\textbf {\bibinfo
  {volume} {112}},\ \bibinfo {pages} {170501} (\bibinfo {year}
  {2014})}\BibitemShut {NoStop}%
\bibitem [{\citenamefont {Poyatos}\ \emph {et~al.}(1996)\citenamefont
  {Poyatos}, \citenamefont {Cirac},\ and\ \citenamefont
  {Zoller}}]{Poyatos_1996}%
  \BibitemOpen
  \bibfield  {author} {\bibinfo {author} {\bibfnamefont {J.~F.}\ \bibnamefont
  {Poyatos}}, \bibinfo {author} {\bibfnamefont {J.~I.}\ \bibnamefont {Cirac}},
  \ and\ \bibinfo {author} {\bibfnamefont {P.}~\bibnamefont {Zoller}},\
  }\href@noop {} {\bibfield  {journal} {\bibinfo  {journal} {Phys. Rev. Lett.}\
  }\textbf {\bibinfo {volume} {77}},\ \bibinfo {pages} {4728} (\bibinfo {year}
  {1996})}\BibitemShut {NoStop}%
\bibitem [{\citenamefont {Verstraete}\ \emph {et~al.}(2009)\citenamefont
  {Verstraete}, \citenamefont {Wolf},\ and\ \citenamefont
  {Ignacio~Cirac}}]{Verstraete_2009}%
  \BibitemOpen
  \bibfield  {author} {\bibinfo {author} {\bibfnamefont {F.}~\bibnamefont
  {Verstraete}}, \bibinfo {author} {\bibfnamefont {M.~M.}\ \bibnamefont
  {Wolf}}, \ and\ \bibinfo {author} {\bibfnamefont {J.}~\bibnamefont
  {Ignacio~Cirac}},\ }\href@noop {} {\bibfield  {journal} {\bibinfo  {journal}
  {Nature Physics}\ }\textbf {\bibinfo {volume} {5}},\ \bibinfo {pages} {633}
  (\bibinfo {year} {2009})}\BibitemShut {NoStop}%
\bibitem [{\citenamefont {Lin}\ \emph {et~al.}(2013)\citenamefont {Lin},
  \citenamefont {Gaebler}, \citenamefont {Reiter}, \citenamefont {Tan},
  \citenamefont {Bowler}, \citenamefont {S{\o}rensen}, \citenamefont
  {Leibfried},\ and\ \citenamefont {Wineland}}]{Wineland}%
  \BibitemOpen
  \bibfield  {author} {\bibinfo {author} {\bibfnamefont {Y.}~\bibnamefont
  {Lin}}, \bibinfo {author} {\bibfnamefont {J.~P.}\ \bibnamefont {Gaebler}},
  \bibinfo {author} {\bibfnamefont {F.}~\bibnamefont {Reiter}}, \bibinfo
  {author} {\bibfnamefont {T.~R.}\ \bibnamefont {Tan}}, \bibinfo {author}
  {\bibfnamefont {R.}~\bibnamefont {Bowler}}, \bibinfo {author} {\bibfnamefont
  {A.~S.}\ \bibnamefont {S{\o}rensen}}, \bibinfo {author} {\bibfnamefont
  {D.}~\bibnamefont {Leibfried}}, \ and\ \bibinfo {author} {\bibfnamefont
  {D.~J.}\ \bibnamefont {Wineland}},\ }\href@noop {} {\bibfield  {journal}
  {\bibinfo  {journal} {Nature}\ }\textbf {\bibinfo {volume} {504}},\ \bibinfo
  {pages} {415} (\bibinfo {year} {2013})}\BibitemShut {NoStop}%
\bibitem [{\citenamefont {Shankar}\ \emph {et~al.}(2013)\citenamefont
  {Shankar}, \citenamefont {Hatridge}, \citenamefont {Leghtas}, \citenamefont
  {Sliwa}, \citenamefont {Narla}, \citenamefont {Vool}, \citenamefont {Girvin},
  \citenamefont {Frunzio}, \citenamefont {Mirrahimi},\ and\ \citenamefont
  {Devoret}}]{Shankar_2013}%
  \BibitemOpen
  \bibfield  {author} {\bibinfo {author} {\bibfnamefont {S.}~\bibnamefont
  {Shankar}}, \bibinfo {author} {\bibfnamefont {M.}~\bibnamefont {Hatridge}},
  \bibinfo {author} {\bibfnamefont {Z.}~\bibnamefont {Leghtas}}, \bibinfo
  {author} {\bibfnamefont {K.~M.}\ \bibnamefont {Sliwa}}, \bibinfo {author}
  {\bibfnamefont {A.}~\bibnamefont {Narla}}, \bibinfo {author} {\bibfnamefont
  {U.}~\bibnamefont {Vool}}, \bibinfo {author} {\bibfnamefont {S.~M.}\
  \bibnamefont {Girvin}}, \bibinfo {author} {\bibfnamefont {L.}~\bibnamefont
  {Frunzio}}, \bibinfo {author} {\bibfnamefont {M.}~\bibnamefont {Mirrahimi}},
  \ and\ \bibinfo {author} {\bibfnamefont {M.~H.}\ \bibnamefont {Devoret}},\
  }\href@noop {} {\bibfield  {journal} {\bibinfo  {journal} {Nature}\ }\textbf
  {\bibinfo {volume} {504}},\ \bibinfo {pages} {419} (\bibinfo {year}
  {2013})}\BibitemShut {NoStop}%
\bibitem [{\citenamefont {Kimchi-Schwartz}\ \emph {et~al.}(2016)\citenamefont
  {Kimchi-Schwartz}, \citenamefont {Martin}, \citenamefont {Flurin},
  \citenamefont {Aron}, \citenamefont {Kulkarni}, \citenamefont {Tureci},\ and\
  \citenamefont {Siddiqi}}]{Kimchi-Schwartz_2016}%
  \BibitemOpen
  \bibfield  {author} {\bibinfo {author} {\bibfnamefont {M.~E.}\ \bibnamefont
  {Kimchi-Schwartz}}, \bibinfo {author} {\bibfnamefont {L.}~\bibnamefont
  {Martin}}, \bibinfo {author} {\bibfnamefont {E.}~\bibnamefont {Flurin}},
  \bibinfo {author} {\bibfnamefont {C.}~\bibnamefont {Aron}}, \bibinfo {author}
  {\bibfnamefont {M.}~\bibnamefont {Kulkarni}}, \bibinfo {author}
  {\bibfnamefont {H.~E.}\ \bibnamefont {Tureci}}, \ and\ \bibinfo {author}
  {\bibfnamefont {I.}~\bibnamefont {Siddiqi}},\ }\href {\doibase
  10.1103/PhysRevLett.116.240503} {\bibfield  {journal} {\bibinfo  {journal}
  {Phys. Rev. Lett.}\ }\textbf {\bibinfo {volume} {116}},\ \bibinfo {pages}
  {240503} (\bibinfo {year} {2016})}\BibitemShut {NoStop}%
\bibitem [{\citenamefont {Bergeal}\ \emph {et~al.}(2010)\citenamefont
  {Bergeal}, \citenamefont {Schackert}, \citenamefont {Metcalfe}, \citenamefont
  {Vijay}, \citenamefont {Manucharyan}, \citenamefont {Frunzio}, \citenamefont
  {Prober}, \citenamefont {Schoelkopf}, \citenamefont {Girvin},\ and\
  \citenamefont {Devoret}}]{Bergeal_2010}%
  \BibitemOpen
  \bibfield  {author} {\bibinfo {author} {\bibfnamefont {N.}~\bibnamefont
  {Bergeal}}, \bibinfo {author} {\bibfnamefont {F.}~\bibnamefont {Schackert}},
  \bibinfo {author} {\bibfnamefont {M.}~\bibnamefont {Metcalfe}}, \bibinfo
  {author} {\bibfnamefont {R.}~\bibnamefont {Vijay}}, \bibinfo {author}
  {\bibfnamefont {V.~E.}\ \bibnamefont {Manucharyan}}, \bibinfo {author}
  {\bibfnamefont {L.}~\bibnamefont {Frunzio}}, \bibinfo {author} {\bibfnamefont
  {D.~E.}\ \bibnamefont {Prober}}, \bibinfo {author} {\bibfnamefont {R.~J.}\
  \bibnamefont {Schoelkopf}}, \bibinfo {author} {\bibfnamefont {S.~M.}\
  \bibnamefont {Girvin}}, \ and\ \bibinfo {author} {\bibfnamefont {M.~H.}\
  \bibnamefont {Devoret}},\ }\href@noop {} {\bibfield  {journal} {\bibinfo
  {journal} {Nature}\ }\textbf {\bibinfo {volume} {465}},\ \bibinfo {pages}
  {64} (\bibinfo {year} {2010})}\BibitemShut {NoStop}%
\bibitem [{\citenamefont {Flurin}\ \emph {et~al.}(2012)\citenamefont {Flurin},
  \citenamefont {Roch}, \citenamefont {Mallet}, \citenamefont {Devoret},\ and\
  \citenamefont {Huard}}]{Flurin_2012}%
  \BibitemOpen
  \bibfield  {author} {\bibinfo {author} {\bibfnamefont {E.}~\bibnamefont
  {Flurin}}, \bibinfo {author} {\bibfnamefont {N.}~\bibnamefont {Roch}},
  \bibinfo {author} {\bibfnamefont {F.}~\bibnamefont {Mallet}}, \bibinfo
  {author} {\bibfnamefont {M.~H.}\ \bibnamefont {Devoret}}, \ and\ \bibinfo
  {author} {\bibfnamefont {B.}~\bibnamefont {Huard}},\ }\href {\doibase
  10.1103/PhysRevLett.109.183901} {\bibfield  {journal} {\bibinfo  {journal}
  {Phys. Rev. Lett.}\ }\textbf {\bibinfo {volume} {109}},\ \bibinfo {pages}
  {183901} (\bibinfo {year} {2012})}\BibitemShut {NoStop}%
\bibitem [{\citenamefont {Flurin}\ \emph {et~al.}(2015)\citenamefont {Flurin},
  \citenamefont {Roch}, \citenamefont {Pillet}, \citenamefont {Mallet},\ and\
  \citenamefont {Huard}}]{Flurin_2015}%
  \BibitemOpen
  \bibfield  {author} {\bibinfo {author} {\bibfnamefont {E.}~\bibnamefont
  {Flurin}}, \bibinfo {author} {\bibfnamefont {N.}~\bibnamefont {Roch}},
  \bibinfo {author} {\bibfnamefont {J.~D.}\ \bibnamefont {Pillet}}, \bibinfo
  {author} {\bibfnamefont {F.}~\bibnamefont {Mallet}}, \ and\ \bibinfo {author}
  {\bibfnamefont {B.}~\bibnamefont {Huard}},\ }\href {\doibase
  10.1103/PhysRevLett.114.090503} {\bibfield  {journal} {\bibinfo  {journal}
  {Phys. Rev. Lett.}\ }\textbf {\bibinfo {volume} {114}},\ \bibinfo {pages}
  {090503} (\bibinfo {year} {2015})}\BibitemShut {NoStop}%
\bibitem [{\citenamefont {Abdo}\ \emph {et~al.}(2013)\citenamefont {Abdo},
  \citenamefont {Sliwa}, \citenamefont {Schackert}, \citenamefont {Bergeal},
  \citenamefont {Hatridge}, \citenamefont {Frunzio}, \citenamefont {Stone},\
  and\ \citenamefont {Devoret}}]{Abdo_2013b}%
  \BibitemOpen
  \bibfield  {author} {\bibinfo {author} {\bibfnamefont {B.}~\bibnamefont
  {Abdo}}, \bibinfo {author} {\bibfnamefont {K.}~\bibnamefont {Sliwa}},
  \bibinfo {author} {\bibfnamefont {F.}~\bibnamefont {Schackert}}, \bibinfo
  {author} {\bibfnamefont {N.}~\bibnamefont {Bergeal}}, \bibinfo {author}
  {\bibfnamefont {M.}~\bibnamefont {Hatridge}}, \bibinfo {author}
  {\bibfnamefont {L.}~\bibnamefont {Frunzio}}, \bibinfo {author} {\bibfnamefont
  {A.~D.}\ \bibnamefont {Stone}}, \ and\ \bibinfo {author} {\bibfnamefont
  {M.}~\bibnamefont {Devoret}},\ }\href {\doibase
  10.1103/PhysRevLett.110.173902} {\bibfield  {journal} {\bibinfo  {journal}
  {Phys. Rev. Lett.}\ }\textbf {\bibinfo {volume} {110}},\ \bibinfo {pages}
  {173902} (\bibinfo {year} {2013})}\BibitemShut {NoStop}%
\bibitem [{\citenamefont {Sirois}\ \emph {et~al.}(2015)\citenamefont {Sirois},
  \citenamefont {Castellanos-Beltran}, \citenamefont {DeFeo}, \citenamefont
  {Ranzani}, \citenamefont {Lecocq}, \citenamefont {Simmonds}, \citenamefont
  {Teufel},\ and\ \citenamefont {Aumentado}}]{Sirois_2015}%
  \BibitemOpen
  \bibfield  {author} {\bibinfo {author} {\bibfnamefont {A.~J.}\ \bibnamefont
  {Sirois}}, \bibinfo {author} {\bibfnamefont {M.~A.}\ \bibnamefont
  {Castellanos-Beltran}}, \bibinfo {author} {\bibfnamefont {M.~P.}\
  \bibnamefont {DeFeo}}, \bibinfo {author} {\bibfnamefont {L.}~\bibnamefont
  {Ranzani}}, \bibinfo {author} {\bibfnamefont {F.}~\bibnamefont {Lecocq}},
  \bibinfo {author} {\bibfnamefont {R.~W.}\ \bibnamefont {Simmonds}}, \bibinfo
  {author} {\bibfnamefont {J.~D.}\ \bibnamefont {Teufel}}, \ and\ \bibinfo
  {author} {\bibfnamefont {J.}~\bibnamefont {Aumentado}},\ }\href {\doibase
  http://dx.doi.org/10.1063/1.4919759} {\bibfield  {journal} {\bibinfo
  {journal} {Appl. Phys. Lett.}\ }\textbf {\bibinfo {volume} {106}},\ \bibinfo
  {pages} {172603} (\bibinfo {year} {2015})}\BibitemShut {NoStop}%
\bibitem [{\citenamefont {Sundaresan}\ \emph {et~al.}(2015)\citenamefont
  {Sundaresan}, \citenamefont {Liu}, \citenamefont {Sadri}, \citenamefont
  {Sz\ifmmode~\mbox{\H{o}}\else \H{o}\fi{}cs}, \citenamefont {Underwood},
  \citenamefont {Malekakhlagh}, \citenamefont {T\"ureci},\ and\ \citenamefont
  {Houck}}]{Sundaresan_2015}%
  \BibitemOpen
  \bibfield  {author} {\bibinfo {author} {\bibfnamefont {N.~M.}\ \bibnamefont
  {Sundaresan}}, \bibinfo {author} {\bibfnamefont {Y.}~\bibnamefont {Liu}},
  \bibinfo {author} {\bibfnamefont {D.}~\bibnamefont {Sadri}}, \bibinfo
  {author} {\bibfnamefont {L.~J.}\ \bibnamefont {Sz\ifmmode~\mbox{\H{o}}\else
  \H{o}\fi{}cs}}, \bibinfo {author} {\bibfnamefont {D.~L.}\ \bibnamefont
  {Underwood}}, \bibinfo {author} {\bibfnamefont {M.}~\bibnamefont
  {Malekakhlagh}}, \bibinfo {author} {\bibfnamefont {H.~E.}\ \bibnamefont
  {T\"ureci}}, \ and\ \bibinfo {author} {\bibfnamefont {A.~A.}\ \bibnamefont
  {Houck}},\ }\href {\doibase 10.1103/PhysRevX.5.021035} {\bibfield  {journal}
  {\bibinfo  {journal} {Phys. Rev. X}\ }\textbf {\bibinfo {volume} {5}},\
  \bibinfo {pages} {021035} (\bibinfo {year} {2015})}\BibitemShut {NoStop}%
\bibitem [{\citenamefont {Bennett}\ \emph {et~al.}(1996)\citenamefont
  {Bennett}, \citenamefont {Bernstein}, \citenamefont {Popescu},\ and\
  \citenamefont {Schumacher}}]{Bennett_1996}%
  \BibitemOpen
  \bibfield  {author} {\bibinfo {author} {\bibfnamefont {C.~H.}\ \bibnamefont
  {Bennett}}, \bibinfo {author} {\bibfnamefont {H.~J.}\ \bibnamefont
  {Bernstein}}, \bibinfo {author} {\bibfnamefont {S.}~\bibnamefont {Popescu}},
  \ and\ \bibinfo {author} {\bibfnamefont {B.}~\bibnamefont {Schumacher}},\
  }\href@noop {} {\bibfield  {journal} {\bibinfo  {journal} {Phys. Rev. A}\
  }\textbf {\bibinfo {volume} {53}},\ \bibinfo {pages} {2046} (\bibinfo {year}
  {1996})}\BibitemShut {NoStop}%
  \bibitem [{\citenamefont {Vollbrecht}\ \emph {et~al.}(2011)\citenamefont
  {Vollbrecht}, \citenamefont {Muschik}, \ and\
  \citenamefont {Cirac}}]{Vollbrecht_2011}%
  \BibitemOpen
  \bibfield  {author} {\bibinfo {author} {\bibfnamefont {K.~G.~H.}\ \bibnamefont
  {Vollbrecht}}, \bibinfo {author} {\bibfnamefont {C.~A.}\ \bibnamefont
  {Muschik}}, \ and\  \bibinfo {author} {\bibfnamefont {J.~I.}~\bibnamefont {Cirac}},\
  }\href@noop {} {\bibfield  {journal} {\bibinfo  {journal} {Phys. Rev. Lett.}\
  }\textbf {\bibinfo {volume} {107}},\ \bibinfo {pages} {120502} (\bibinfo {year}
  {2011})}\BibitemShut {NoStop}%
\bibitem [{\citenamefont {Kraus}\ and\ \citenamefont
  {Cirac}(2004)}]{Kraus_2004}%
  \BibitemOpen
  \bibfield  {author} {\bibinfo {author} {\bibfnamefont {B.}~\bibnamefont
  {Kraus}}\ and\ \bibinfo {author} {\bibfnamefont {J.~I.}\ \bibnamefont
  {Cirac}},\ }\href {\doibase 10.1103/PhysRevLett.92.013602} {\bibfield
  {journal} {\bibinfo  {journal} {Phys. Rev. Lett.}\ }\textbf {\bibinfo
  {volume} {92}},\ \bibinfo {pages} {013602} (\bibinfo {year}
  {2004})}\BibitemShut {NoStop}%
\bibitem [{\citenamefont {Aron}\ \emph {et~al.}(2014)\citenamefont {Aron},
  \citenamefont {Kulkarni},\ and\ \citenamefont {T\"ureci}}]{Aron_2014}%
  \BibitemOpen
  \bibfield  {author} {\bibinfo {author} {\bibfnamefont {C.}~\bibnamefont
  {Aron}}, \bibinfo {author} {\bibfnamefont {M.}~\bibnamefont {Kulkarni}}, \
  and\ \bibinfo {author} {\bibfnamefont {H.~E.}\ \bibnamefont {T\"ureci}},\
  }\href {\doibase 10.1103/PhysRevA.90.062305} {\bibfield  {journal} {\bibinfo
  {journal} {Phys. Rev. A}\ }\textbf {\bibinfo {volume} {90}},\ \bibinfo
  {pages} {062305} (\bibinfo {year} {2014})}\BibitemShut {NoStop}%
\bibitem [{\citenamefont {Motzoi}\ \emph {et~al.}(2016)\citenamefont {Motzoi},
  \citenamefont {Halperin}, \citenamefont {Wang}, \citenamefont {Whaley},\ and\
  \citenamefont {Schirmer}}]{motzoi_2016}%
  \BibitemOpen
  \bibfield  {author} {\bibinfo {author} {\bibfnamefont {F.}~\bibnamefont
  {Motzoi}}, \bibinfo {author} {\bibfnamefont {E.}~\bibnamefont {Halperin}},
  \bibinfo {author} {\bibfnamefont {X.}~\bibnamefont {Wang}}, \bibinfo {author}
  {\bibfnamefont {K.~B.}\ \bibnamefont {Whaley}}, \ and\ \bibinfo {author}
  {\bibfnamefont {S.}~\bibnamefont {Schirmer}},\ }\href {\doibase
  10.1103/PhysRevA.94.032313} {\bibfield  {journal} {\bibinfo  {journal} {Phys.
  Rev. A}\ }\textbf {\bibinfo {volume} {94}},\ \bibinfo {pages} {032313}
  (\bibinfo {year} {2016})}\BibitemShut {NoStop}%
\bibitem [{\citenamefont {Shi}\ and\ \citenamefont {Nurdin}(2015)}]{Shi_2015}%
  \BibitemOpen
  \bibfield  {author} {\bibinfo {author} {\bibfnamefont {Z.}~\bibnamefont
  {Shi}}\ and\ \bibinfo {author} {\bibfnamefont {H.~I.}\ \bibnamefont
  {Nurdin}},\ }\href {\doibase 10.1007/s11128-014-0845-4} {\bibfield  {journal}
  {\bibinfo  {journal} {Quant. Inf. Process.}\ }\textbf {\bibinfo {volume}
  {14}},\ \bibinfo {pages} {337} (\bibinfo {year} {2015})}\BibitemShut
  {NoStop}%
\bibitem [{\citenamefont {Wang}\ and\ \citenamefont {Clerk}(2012)}]{Wang_2012}%
  \BibitemOpen
  \bibfield  {author} {\bibinfo {author} {\bibfnamefont {Y.-D.}\ \bibnamefont
  {Wang}}\ and\ \bibinfo {author} {\bibfnamefont {A.~A.}\ \bibnamefont
  {Clerk}},\ }\href {\doibase 10.1103/PhysRevLett.108.153603} {\bibfield
  {journal} {\bibinfo  {journal} {Phys. Rev. Lett.}\ }\textbf {\bibinfo
  {volume} {108}},\ \bibinfo {pages} {153603} (\bibinfo {year}
  {2012})}\BibitemShut {NoStop}%
\bibitem [{Note1()}]{Note1}%
  \BibitemOpen
  \bibinfo {note} {The ratio of standard deviations between the anti-squeezed
  and squeezed quadratures is $e^{2r}=10^{r_\protect \mathrm
  {dB}/10}$.}\BibitemShut {Stop}%
\bibitem [{SM()}]{SM}%
  \BibitemOpen
  \href@noop {} {}\bibinfo {note} {See Supplemental Material.}\BibitemShut
  {Stop}%
\bibitem [{\citenamefont {Wootters}(2001)}]{Wootters_2001}%
  \BibitemOpen
  \bibfield  {author} {\bibinfo {author} {\bibfnamefont {W.~K.}\ \bibnamefont
  {Wootters}},\ }\href@noop {} {\bibfield  {journal} {\bibinfo  {journal}
  {Quantum Info. Comput.}\ }\textbf {\bibinfo {volume} {1}},\ \bibinfo {pages}
  {27} (\bibinfo {year} {2001})}\BibitemShut {NoStop}%
\bibitem [{\citenamefont {Roy}\ and\ \citenamefont {Devoret}(2016)}]{Roy_2016}%
  \BibitemOpen
  \bibfield  {author} {\bibinfo {author} {\bibfnamefont {A.}~\bibnamefont
  {Roy}}\ and\ \bibinfo {author} {\bibfnamefont {M.}~\bibnamefont {Devoret}},\
  }\href@noop {} {\bibfield  {journal} {\bibinfo  {journal} {Comptes Rendus
  Physique}\ }\textbf {\bibinfo {volume} {17}},\ \bibinfo {pages} {740 }
  (\bibinfo {year} {2016})}\BibitemShut {NoStop}%
\bibitem [{\citenamefont {Leghtas}\ \emph {et~al.}(2013)\citenamefont
  {Leghtas}, \citenamefont {Vool}, \citenamefont {Shankar}, \citenamefont
  {Hatridge}, \citenamefont {Girvin}, \citenamefont {Devoret},\ and\
  \citenamefont {Mirrahimi}}]{Leghtas_2013}%
  \BibitemOpen
  \bibfield  {author} {\bibinfo {author} {\bibfnamefont {Z.}~\bibnamefont
  {Leghtas}}, \bibinfo {author} {\bibfnamefont {U.}~\bibnamefont {Vool}},
  \bibinfo {author} {\bibfnamefont {S.}~\bibnamefont {Shankar}}, \bibinfo
  {author} {\bibfnamefont {M.}~\bibnamefont {Hatridge}}, \bibinfo {author}
  {\bibfnamefont {S.~M.}\ \bibnamefont {Girvin}}, \bibinfo {author}
  {\bibfnamefont {M.~H.}\ \bibnamefont {Devoret}}, \ and\ \bibinfo {author}
  {\bibfnamefont {M.}~\bibnamefont {Mirrahimi}},\ }\href {\doibase
  10.1103/PhysRevA.88.023849} {\bibfield  {journal} {\bibinfo  {journal} {Phys.
  Rev. A}\ }\textbf {\bibinfo {volume} {88}},\ \bibinfo {pages} {023849}
  (\bibinfo {year} {2013})}\BibitemShut {NoStop}%
\bibitem [{\citenamefont {Liu}\ \emph {et~al.}(2016)\citenamefont {Liu},
  \citenamefont {Shankar}, \citenamefont {Ofek}, \citenamefont {Hatridge},
  \citenamefont {Narla}, \citenamefont {Sliwa}, \citenamefont {Frunzio},
  \citenamefont {Schoelkopf},\ and\ \citenamefont {Devoret}}]{Liu_2016}%
  \BibitemOpen
  \bibfield  {author} {\bibinfo {author} {\bibfnamefont {Y.}~\bibnamefont
  {Liu}}, \bibinfo {author} {\bibfnamefont {S.}~\bibnamefont {Shankar}},
  \bibinfo {author} {\bibfnamefont {N.}~\bibnamefont {Ofek}}, \bibinfo {author}
  {\bibfnamefont {M.}~\bibnamefont {Hatridge}}, \bibinfo {author}
  {\bibfnamefont {A.}~\bibnamefont {Narla}}, \bibinfo {author} {\bibfnamefont
  {K.~M.}\ \bibnamefont {Sliwa}}, \bibinfo {author} {\bibfnamefont
  {L.}~\bibnamefont {Frunzio}}, \bibinfo {author} {\bibfnamefont {R.~J.}\
  \bibnamefont {Schoelkopf}}, \ and\ \bibinfo {author} {\bibfnamefont {M.~H.}\
  \bibnamefont {Devoret}},\ }\href {\doibase 10.1103/PhysRevX.6.011022}
  {\bibfield  {journal} {\bibinfo  {journal} {Phys. Rev. X}\ }\textbf {\bibinfo
  {volume} {6}},\ \bibinfo {pages} {011022} (\bibinfo {year}
  {2016})}\BibitemShut {NoStop}%
\end{thebibliography}
\end{document}